\documentclass[sigconf]{acmart}

\AtBeginDocument{%
  }

\setcopyright{acmlicensed}
\copyrightyear{2018}
\acmYear{2018}
\acmDOI{XXXXXXX.XXXXXXX}

\acmConference[SIGCSE '24]{Special Interest Group on Computer Science Education Virtual 2024}{December 05-07,
  2024}{Virtual}
\acmISBN{978-1-4503-XXXX-X/18/06}

\usepackage{float} 
\usepackage{booktabs}
\usepackage{array}
\begin{document}

\title{CourseAssist: Pedagogically Appropriate AI Tutor for Computer Science Education}

\author{Ty Feng}
\email{ty@courseassistai.com}
\orcid{0000-0002-7428-3141}
\affiliation{%
  \institution{CourseAssist, Inc.}
  \city{Davis}
  \state{California}
  \country{USA}
}

\author{Sa Liu}
\email{sarah@courseassistai.com}
\affiliation{%
  \institution{CourseAssist, Inc.}
  \city{Davis}
  \state{California}
  \country{USA}
}

\author{Dipak Ghosal}
\email{dghosal@ucdavis.edu}
\affiliation{%
  \institution{University of California, Davis}
  \city{Davis}
  \state{California}
  \country{USA}
}

\renewcommand{\shortauthors}{Feng et al.}

\begin{abstract}
  The growing enrollments in computer science courses and increase in class sizes necessitate scalable, automated tutoring solutions to adequately support student learning. While Large Language Models (LLMs) like GPT-4 have demonstrated potential in assisting students through question-answering, educators express concerns over student overreliance, miscomprehension of generated code, and the risk of inaccurate answers. Rather than banning these tools outright, we advocate for a constructive approach that harnesses the capabilities of AI while mitigating potential risks. This poster introduces CourseAssist\footnote{CourseAssist can be accessed at \url{https://courseassistai.com}}, a novel LLM-based tutoring system tailored for computer science education. Unlike generic LLM systems, CourseAssist uses retrieval-augmented generation, user intent classification, and question decomposition to align AI responses with specific course materials and learning objectives, thereby ensuring pedagogical appropriateness of LLMs in educational settings. We evaluated CourseAssist against a baseline of GPT-4 using a dataset of 50 question-answer pairs from a programming languages course, focusing on the criteria of usefulness, accuracy, and pedagogical appropriateness. Evaluation results show that CourseAssist significantly outperforms the baseline, demonstrating its potential to serve as an effective learning assistant. We have also deployed CourseAssist in 6 computer science courses at a large public R1 research university reaching over 500 students. Interviews with 20 student users show that CourseAssist improves computer science instruction by increasing the accessibility of course-specific tutoring help and shortening the feedback loop on their programming assignments. Future work will include extensive pilot testing at more universities and exploring better collaborative relationships between students, educators, and AI that improve computer science learning experiences.
\end{abstract}



\maketitle

\section{Introduction}
With increasing enrollments in computer science and expanding class sizes, there is an acute need for scalable, automated tutoring solutions that can provide students with instant and adequate support. Recent advances in large language models (LLMs) have shown their potential in helping students learn as a question-answering tutor, but it is difficult to maintain academic integrity when students use LLMs to generate code solutions for assignments. While LLMs are already used by students for programming help, we are still in the early stage of exploring how LLMs may be used to best support learning and discovery by students and how to minimize the potential risks and harms from LLMs in computer science education. When students are using LLMs without any guardrails or oversight by educators, the use of LLMs may result in overreliance on LLMs for code authorship, lack of comprehension of LLM code, and learning misleading or suboptimal code solutions that are inconsistent with learning goals. These challenges may hinder students from learning computer science and becoming more adept at programming.  

While recent studies have explored the role of LLMs in programming education~\cite{wieser, Hellas, kazemitabaar2023novices, gpt-pass-assessments, patterns-help-seeking-llm}, most studies have only studied generic LLM systems like ChatGPT without modifying LLM behavior or customizing it for computer science education. As an emerging technology, there is a lot of potential to explore different ways to improve and enhance LLM capabilities and utility for computer science education. One of the issues we are interested in is how LLMs can be steered and customized to align their responses with course-specific learning goals. We observed that LLMs can generate responses that conflict with learning objectives set by course instructors, and currently there is little option for instructors to control the behaviors of LLMs and how students should use them. Instead of attempting to ban the use of LLM systems like ChatGPT and Copilot in classrooms, we propose a harm-reductionist approach to address some of the challenges and issues with students using LLMs for computer science courses.

There are several LLM-powered tools deployed in real classroom settings, such as CodeHelp~\cite{CodeHelp}, which responds to semi-structured student queries without providing exact solution code regardless of how a student writes the query. To use this tool, however, students would need to write structured descriptions of their problem in order for it to understand the user's intent, which can be cumbersome for the user. In an effort to provide a better solution, we developed CourseAssist, a LLM-based tutoring system that produces responses more aligned with the learning goals of a specific computer science course, by building custom query pipelines to better handle various user intents. 

In this work, we evaluate CourseAssist on whether it generates more educationally appropriate responses for a given computer science course compared to the baseline ChatGPT. Specifically, we propose that pedagogical appropriateness should be considered when evaluating LLM-based tutoring systems. Pedagogical appropriateness means whether the system promotes learning on the student's part. One common method that human teachers use is giving hints to guide students to work through problems, rather than showing direct answers to every question. Thus, it is important for LLM-based tutoring systems to emulate this behavior and ensure LLM-generated responses are appropriate for learning. 
Our main contributions in this work are:
\begin{itemize}
    \item We introduce a LLM-based question answering system that provides course-specific guidance and hints using retrieval-augmented generation and user intent classification to ensure pedagogical appropriateness.  
    \item We present a human evaluation method that considers usefulness, accuracy, and pedagogical appropriateness as the evaluation criteria. 
    \item We investigate the applicability of CourseAssist to act as a tutor in computer science courses by evaluating its responses on a Piazza QA dataset from a real-world computer science course and compare it with the baseline ChatGPT 4. 
\end{itemize}

\section{Method}
\subsection{CourseAssist}
The primary goal of this study is to evaluate whether LLMs can be used in a controllable manner for computer science education. We developed CourseAssist, an AI tutoring system, by augmenting GPT-4 with user intent classification, question decomposition, and retrieval of course content. We developed a retrieval engine that can decide when and what to retrieve based on the user intent, so that retrievals are more accurate for computer science students. 

\section{Dataset}
For the experiments in this study, we sampled 50 question answer pairs from 185 Piazza posts from a programming languages course at a large public university. We rewrote all of the student questions to preserve their semantic meaning while improving the clarity and conciseness. To ensure data quality of the answers, we used the instructor or TA's answer for each question and discarded answers by other students. None of the question answer pairs have any personal information, and the data is fully anonymized. Of the 50 questions, 33 are homework related questions, and 17 are conceptual questions about lecture materials. The dataset and evaluation rubric can be found here: \url{https://github.com/ytyfeng/CourseAssist-Data}. 

\section{Results}
To evaluate the system, we compare CourseAssist answers and answers generated by the baseline GPT-4 to the 50 Piazza questions. Previous studies \cite{aita-2023} have shown that automatic metrics, such as BLEU, Meteor, Rouge-L, and Cider, are not effective at evaluating usefulness of generated text. Therefore, we conducted a human evaluation to evaluate usefulness, accuracy, and pedagogical appropriateness of the generated responses. On each evaluation criterion, we asked an instructor of the course to score the generated response on a three-point scale of 0, 0.5, and 1. 

All experiments were run two times to account for variability in LLM responses, and the evaluation scores are the average of the two runs. The results showed that CourseAssist outperformed the baseline GPT 4 in all of the evaluation criteria. Not only is CourseAssist more useful and accurate than GPT 4, every answer it generated is pedagogically appropriate due to the user intent classification and post-processing steps. 

\begin{table}[H]
    \centering
    \caption{Human evaluation scores (0: poor; 0.5: fair; 1: good) of CourseAssist and GPT 4 based on usefulness, accuracy, and pedagogical appropriateness on 50 QA pairs}
    \begin{tabular}{>{\bfseries}p{2.5cm}|>{\centering\arraybackslash}p{2.4cm}|>{\centering\arraybackslash}p{2.4cm}}
        \toprule
        \textbf{Criteria} & \textbf{CourseAssist} & \textbf{GPT-4} \\
        \midrule
        Usefulness       & \textbf{0.91} & 0.83 \\
        Accuracy         & \textbf{0.92} & 0.71 \\
        Appropriateness  & \textbf{1.00} & 0.82 \\
        Average          & \textbf{0.94} & 0.79 \\
        \bottomrule
    \end{tabular}
    \label{tab:results}
\end{table}

\bibliographystyle{ACM-Reference-Format}
\bibliography{sample-base}

\end{document}